\documentclass[a4paper,11pt]{article}
\usepackage{setspace}
\usepackage{graphicx}

\usepackage{pos}
\newcommand{\N}{{\mathcal N}}
\newcommand{\beq}{\begin{equation}}
\newcommand{\eeq}{\end{equation}}
\newcommand{\bea}{\begin{eqnarray}}
\newcommand{\eea}{\end{eqnarray}}
\usepackage[capitalise]{cleveref}

\definecolor{orangeQCD}{rgb}{0.9725490196078431,0.6274509803921569,0.49411764705882355}
\definecolor{colZN}{rgb}{0.968735,0.465621,0.212103}
\definecolor{greenAstro}{rgb}{0.000000,0.427451,0.172549}
\definecolor{greenAstroL}{rgb}{0.0, 0.66, 0.42}
\definecolor{blueEDM}{rgb}{0.031373,0.317647,0.611765}
\definecolor{BlueC0}{rgb}{0.000000,0.447059,0.698039}
\definecolor{GreenC1}{rgb}{0.000000,0.619608,0.450980}
\definecolor{OrangeC2}{rgb}{0.835294,0.368627,0.000000}

\definecolor{colFinDen}{rgb}{0.894118,0.529412,0.317647}
\definecolor{colSNSup}{rgb}{0.137255,0.545098,0.270588}
\definecolor{colnEDM}{rgb}{0.031373,0.317647,0.611765}
\definecolor{colLab}{rgb}{0.066667,0.317647,0.172549}
\definecolor{colAstro}{rgb}{0.000000,0.427451,0.172549}
\definecolor{colAstroL}{rgb}{0.0, 0.66, 0.42}
\definecolor{colZN}{rgb}{0.968735,0.465621,0.212103}
\definecolor{purpleDMbands}{rgb}{0.623529,0.141176,0.384314}

\title{An exceptionally light axion: Strong CP and Dark Matter}

\author*[a]{Pablo Quilez}
\author[a,b,c]{Luca Di Luzio}
\author[d,e]{Belen Gavela}
\author[a]{Andreas Ringwald}
\affiliation[a]{Deutsches Elektronen-Synchrotron DESY, Notkestr. 85, 22607 Hamburg, Germany}

\affiliation[b]{Dipartimento di Fisica e Astronomia ‘G. Galilei’, Universit\`a di Padova, Italy}
\affiliation[c]{Istituto Nazionale Fisica Nucleare, Sezione di Padova, Italy}
\affiliation[d]{Departamento de F\'isica Te\'orica, Universidad Aut\'onoma de Madrid, 
Cantoblanco, 28049, Madrid, Spain}
\affiliation[e]{Instituto de F\'isica Te\'orica, IFT-UAM/CSIC,
Cantoblanco, 28049, Madrid, Spain}

\emailAdd{pablo.quilez@desy.de}

\abstract{We explore whether the axion which solves the strong CP problem can naturally be much lighter than the canonical QCD axion. The $Z_\N$ symmetry proposed by Hook, with $\N$ mirror and degenerate worlds coexisting in Nature and linked by the axion field, is considered and the associated phenomenology is studied in detail.
On a second step, we show that dark matter can be accounted for by this extremely light axion. This includes the first proposal of a “fuzzy dark matter” QCD axion. A novel misalignment mechanism occurs – trapped
misalignment– due to the peculiar temperature dependence of the $Z_\N$ axion potential, which in some cases can also dynamically source the recently proposed kinetic misalignment mechanism.
The resulting universal enhancement of all axion interactions relative to those of the canonical QCD axion has a strong impact on the prospects of ALP experiments such as ALPS II, IAXO and many others. For instance, even Phase I of Casper Electric could discover this axion.}

\FullConference{%
  *** The European Physical Society Conference on High Energy Physics (EPS-HEP2021), ***\\
  *** 26-30 July 2021 ***\\
  *** Online conference, jointly organized by Universität Hamburg and the research center DESY ***
}

\begin{document}

\begin{flushright}
    DESY 21-180\\
\end{flushright}
\maketitle

\section{Motivation and canonical axion mass}
The QCD axion is one of the most motivated scenarios beyond the Standard Model. This simple extension not only explains in an elegant way the absence of CP violation in the strong interactions~\cite{Peccei:1977hh,Peccei1977} but could also account for the Dark matter abundance~\cite{Preskill:1982cy,Abbott:1982af,Dine:1982ah}. Furthermore, the QCD axion is a highly predictive scenario since a lot of the properties of this pseudoscalar stem from its pseudo-Goldstone boson nature: both the axion mass 
and  the couplings to ordinary matter scale as $1/f_a$, where $f_a$ is the 
axion decay constant, denoting the scale at which the Peccei-Quinn (PQ) symmetry $U(1)_{\rm PQ}$ is spontaneously broken.

At the heart of the axion solution to the strong CP problem lies the fact that the QCD anomaly is the only source of explicit PQ breaking. As a byproduct, within the so-called canonical axions the $m_a$-$f_a$ relation is fixed by QCD~\cite{Weinberg:1977ma,Wilczek1978},
\begin{align}
m_{a}^{\rm QCD} \simeq \frac{f_{\pi} m_{\pi}}{f_{a}} \frac{m_{u} m_{d}}{\sqrt{m_{u}+m_{d}}}\,,
\label{Eq: axion mass}
\end{align}
where $m_\pi, f_\pi, m_u$ and $m_d$ denote respectively  the pion mass, its decay constant, and the up  and down quark masses. The strength of the axion couplings to Standard Model (SM) fields is instead model-dependent: it varies with the matter content of the UV complete axion model. 

 In recent years there have been many attempts to enlarge the canonical QCD 
axion window, by considering UV  completions of the axion effective Lagrangian 
which depart from the minimal invisible axion constructions \cite{Zhitnitsky:1980tq,Dine:1981rt,Kim:1979if,Shifman:1979if}. Most approaches actually focused on the possibility of modifying 
the Wilson coefficient of  specific axion-SM effective operators (see Ref.~\cite{Agrawal:2021dbo} and Refs. therein). A most intriguing possibility consists on departing from the $m_a$-$f_a$ relation in Eq.~(\ref{Eq: axion mass}). Indeed axions which are heavier than the canonical QCD axion have been explored since long and  revived in the last years (see e.g. Refs.~\cite{rubakov:1997vp,Gaillard:2018xgk,Csaki:2019vte}).
  In contrast, solutions to the strong CP problem with lighter axions were uncharted territory until very recently.

The goal of this work is to study in detail and determine the phenomenological implications of a freshly proposed dynamical --and technically natural-- scenario, which solves the strong CP problem with an axion much lighter than the canonical QCD one~\cite{Hook:2018jle,ZNCPpaper}. Next, we show that dark matter
can be accounted for by this extremely light axion and that features a novel production mechanism in the early Universe: the \emph{trapped misalignment mechanism}~\cite{DiLuzio:2021gos}.

\section{The non-linearly realized $Z_\N$ axion}
Let's assume that Nature is endowed with a $Z_\N$ symmetry under which $\N$ copies of the SM are interchanged and which is non-linearly realized by the axion field~\cite{Hook:2018jle,ZNCPpaper},
 \begin{align}
  Z_\N:\quad &\text{SM}_{k} \longrightarrow \text{SM}_{k+1\,(\text{mod} \,\N)} \label{mirror-charges}\\
       & a \longrightarrow a + \frac{2\pi k}{\N} f_a\,.
       \label{axion-detuned-charge}
\end{align}
Given this symmetry, $\N$ mirror and degenerate worlds linked by the axion field would coexist with the same coupling strengths as in the SM, with the exception of the 
effective $\theta_k$-parameters: for each copy $k$ the effective $\theta$-value is shifted by $2\pi/\N$ with respect to that in the 
neighboring $k-1$ sector. Thus the total potential for the axion is given by the sum of all the shifted contributions,
\begin{align}
V_\N(\theta_a)=-m_{\pi}^{2} f_{\pi}^{2}\sum_{k=0}^{\N-1}  \sqrt{1-\frac{4 m_{u} m_{d}}{\left(m_{u}+m_{d}\right)^{2}} \sin ^{2}\left(\frac{\theta_a}{2}+\frac{\pi k}{\N}\right)}\, ,
\label{Eq:Vsmilga ZN}
\end{align}
where $\theta_a \equiv a / f_a$ is the angular axion field.
Strikingly, the resulting axion is exponentially lighter than the canonical QCD axion in Eq.~(\ref{Eq: axion mass}), because the non-perturbative contributions to its potential from the $\N$ degenerate QCD groups  conspire by symmetry to suppress each other~\cite{Hook:2018jle,ZNCPpaper}.
  Indeed, it has been shown that this kind of $Z_\N$ symmetric potentials have interesting mathematical properties\footnote{See Ref.~\cite{Das:2020arz} for a generalization of the mechanism for non-abelian discrete symmetries.} in the large $\N$ limit and for this case the total axion potential is given in all generality by a compact analytical formula~\cite{ZNCPpaper},
  \begin{align}
   \label{Eq: fourier potential large N hyper-compact}
V_{\mathcal{N}}\left(\theta_a\right)
  \simeq - \frac{m_a^2 f_a^2}{\N^2} \,\cos (\N\theta_a)\,, \qquad 
  m_a^2 \, f^2_a  \simeq  \frac{m_{\pi}^{2} f_{\pi}^{2}}{\sqrt{\pi}} \,\sqrt{\frac{1-z}{1+z}} \,\N^{3/2} \,z^\N\,,
  \end{align}
  where the exponential suppression of the $Z_\N$ axion mass squared ($\propto 2^{-\N}$)  in comparison to the canonical case  $\left(m_a^{\rm QCD}\right)^2$ in Eq.~(\ref{Eq: axion mass}) is controlled by the ratio of light quark masses $z\equiv m_u/m_d\simeq 1/2$.

  The solution to the strong CP problem of this $Z_\N$ scenario required  $\N$ to be odd. 
 Overall, the $\sim 10$ orders of magnitude tuning require by the SM strong CP problem is traded by a $1/\N$ adjustment, where $\N$ could be as low as $\N=3$. 
 \begin{figure}[!h]
\centering
\includegraphics[width=0.5\textwidth]{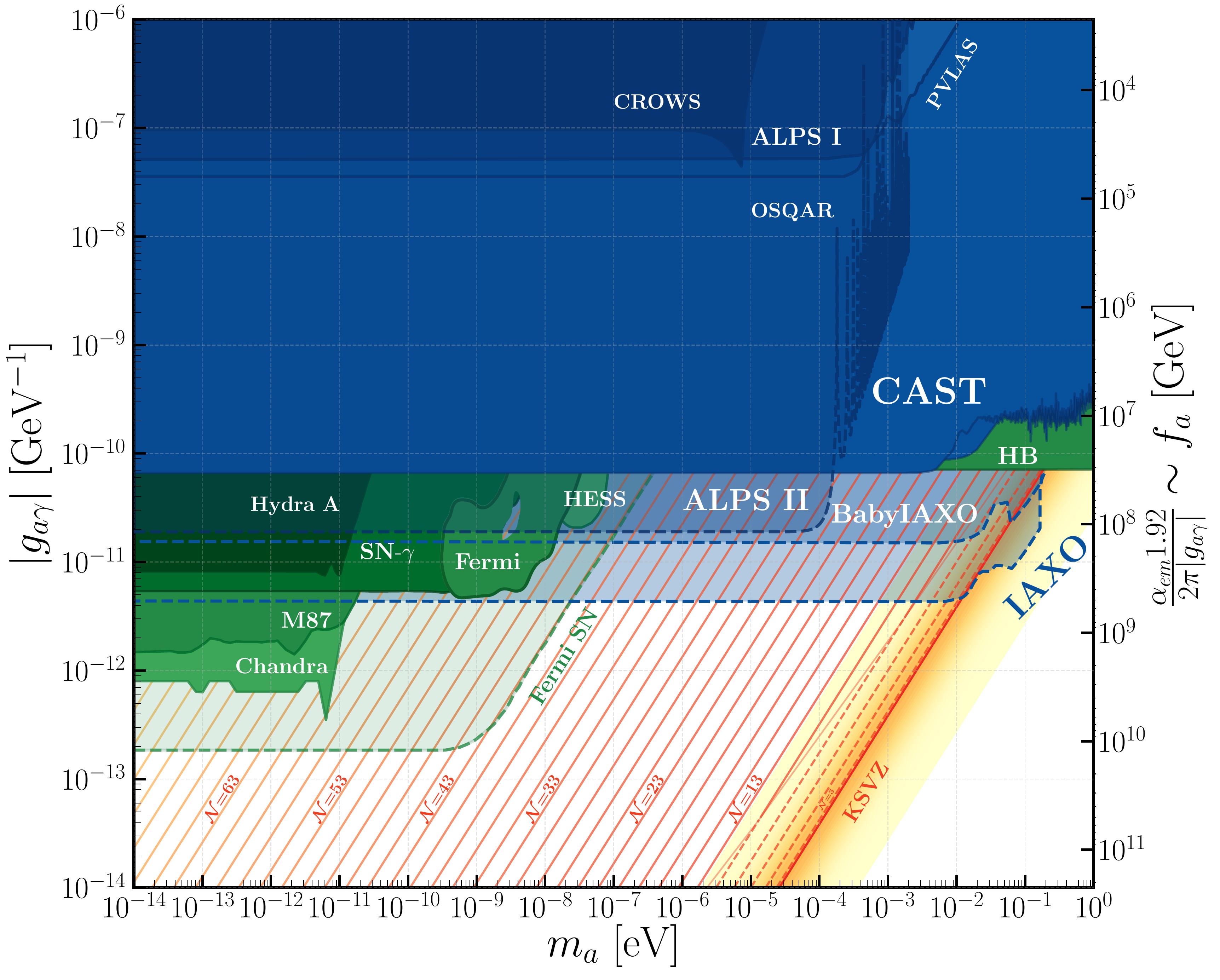}
\includegraphics[width=0.45\textwidth]{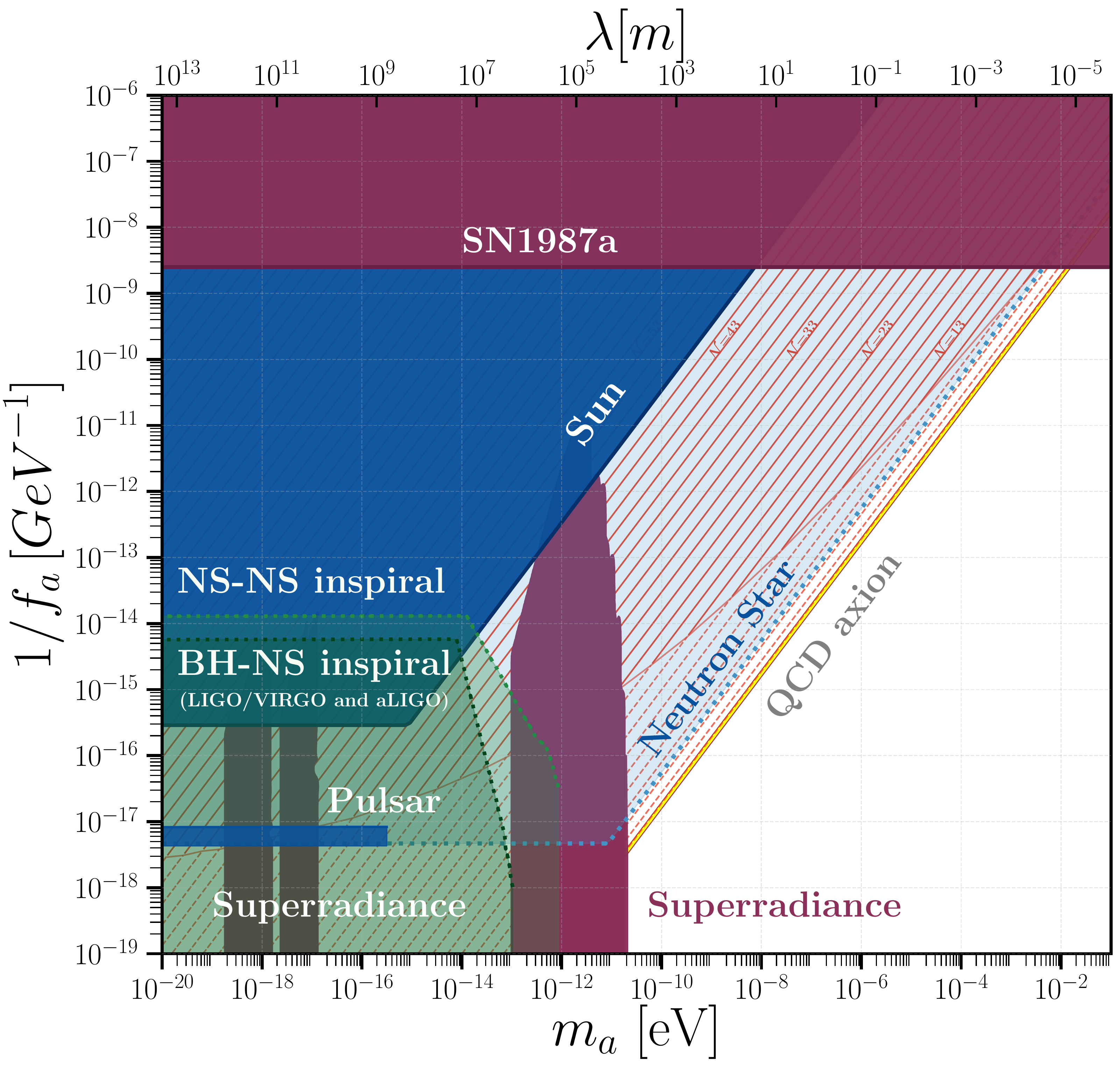}
\caption{\small  Limits on the axion-photon coupling (left) and on the inverse of the axion decay constant (right) as a function of the axion mass. The  {\color{OrangeC2} \bf orange}  oblique lines represent the theoretical prediction for the $Z_\N$ axion case (assuming vanishing ``bare'' coupling to photons)
 for different $\N$. 
See Refs.~\cite{ciaran_o_hare_2020_3932430,ZNCPpaper} for details.}
\label{fig:Photon coupling no DM}       
\end{figure}

The crucial properties of such a light axion are generic and do not depend on the details of the putative UV completion. An important byproduct of this construction  is  an enhancement of all axion interactions which is {\it universal}, that is, model-independent and equal for all axion couplings, at fixed $m_a$, see Fig.~(\ref{fig:Photon coupling no DM}).  The detailed exploration  of the $Z_\N$ paradigm and of the phenomenological constraints which do not require the axion to account for DM can be found in Ref.~\cite{ZNCPpaper}. 
It is particularly enticing that experiments set {\it a priori} to hunt only for ALPs may in fact be targeting solutions to the strong CP problem.  For instance,  
ALPS II is shown to be able to probe the $Z_\N$ scenario here discussed down to 
$\N \sim 25$ for a large enough axion-photon coupling, while IAXO and BabyIAXO may test the whole $\N$ landscape for  values of that coupling even smaller, see Fig.~(\ref{fig:Photon coupling no DM}). 

Furthermore, highly dense stellar bodies allow one to set even stronger bounds in wide regions of the parameter space. These exciting limits have an added value: they avoid model-dependent assumptions about the  axion couplings to SM particles, because they rely exclusively on the anomalous axion-gluon interaction needed to solve to the strong CP problem. A dense medium of ordinary matter is a background that breaks the $Z_\N$ symmetry. This hampers the symmetry-induced cancellations in the total axion potential:
 the axion becomes heavier inside dense media {\it and} the minimum of the potential is located at $\theta_a=\pi$. The corresponding bounds from present solar data, together with projections with neutron stars are shown in Fig.~(\ref{fig:Photon coupling no DM}) (right). Moreover, gravitational wave data from NS-NS and BH-NS mergers by LIGO/VIRGO and Advanced LIGO will allow to further probe this setup \cite{Hook:2017psm,Zhang:2021mks}.

 For the sake of illustration, we have also developed two examples of UV completed models featuring this mechanism~\cite{ZNCPpaper}. Specially interesting is the $Z_\N$ KSVZ model, which is shown to enjoy an improved PQ quality behaviour in large regions of the parameter space, depicted in solid orange lines in Fig.~(\ref{fig:Photon coupling no DM}), see e.g. Refs.~\cite{Redi:2016esr,Gavela:2018paw} for alternative solutions to the PQ quality problem. 

\section{$Z_\N$ axion dark matter} 
\label{sec:axion_dark_matter}

The evolution of the $Z_\N$ axion field 
and its contribution to the DM relic abundance  
departs drastically from the standard case~\cite{DiLuzio:2021gos}.   
The cosmological impact of hypothetical parallel ``mirror''  worlds has been studied at length in the literature (for a review, see e.g. Ref.~\cite{Berezhiani:2003xm}). 
Crucially, the constraints on the number of effective relativistic species $N_{\text{eff}}$ imply that the mirror copies of the SM must be less populated\footnote{Mechanisms that source this world-asymmetric initial temperatures while preserving the $Z_\N$ symmetry may arise naturally in the cosmological 
 evolution~\cite{Kolb:1985bf,Dvali:2019ewm}. }  --cooler-- than the ordinary SM world. 
\begin{figure}[!h]
\centering
\includegraphics[width=0.9\textwidth]{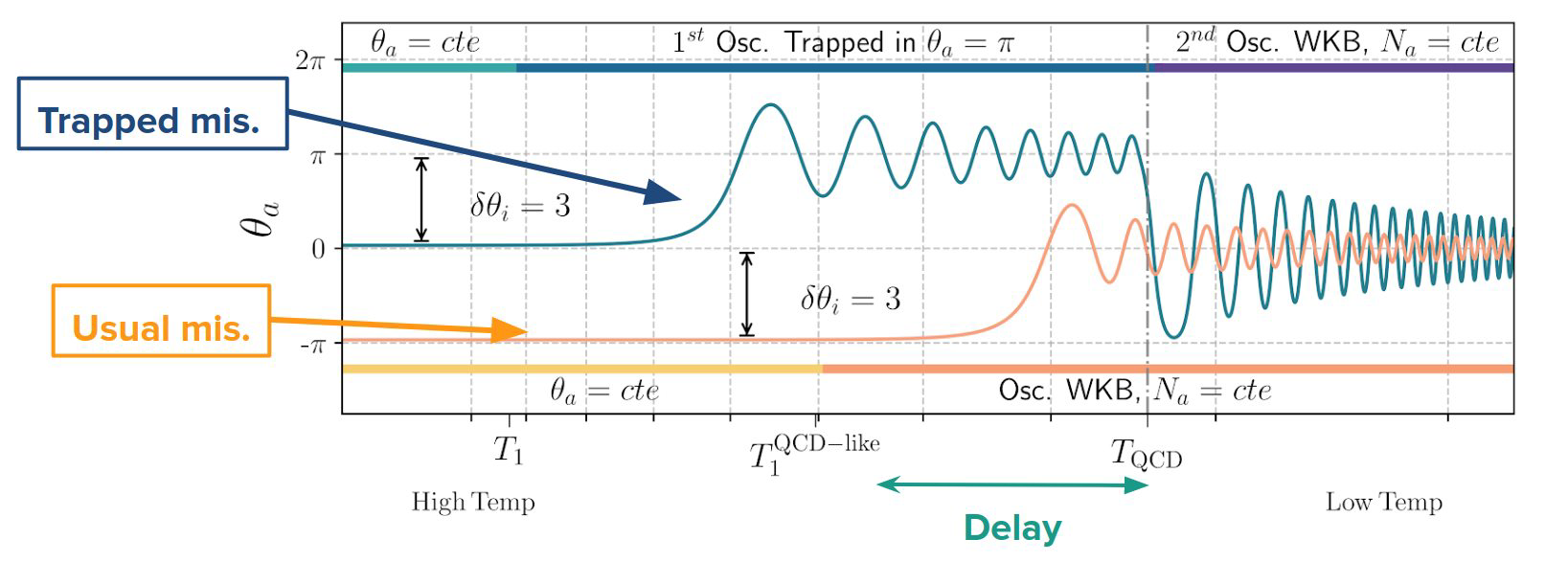}
\caption{\small  Comparison of the evolution of the axion field in the trapped case vs standard misalignment}
\label{fig:Trapped}       
\end{figure}
 As a consequence of this temperature asymmetry among the worlds, and similarly to the previously described finite density effects, the temperature dependence of the $Z_{\N}$ axion potential presents some particular features that modify the production of DM axions in terms of the misalignment mechanism. The scenario results in a novel type of misalignment, with a large value of the misalignment angle. 
 In particular, the relic density is enhanced because the axion field undergoes two stages of oscillations, 
  that are separated by 
  an unavoidable and 
  drastic --non-adiabatic-- modification of the potential. The axion field
  is first \emph{trapped} in the wrong minimum (with $\theta=\pi$), which effectively delays the onset of the true oscillations and thus enhances the DM density.
We will call this new production mechanism \emph{trapped misalignment}, see Fig.~(\ref{fig:Trapped}).
Furthermore, in some regions of the parameter space,  trapped misalignment will automatically source the recently proposed kinetic misalignment mechanism~\cite{Co:2019jts}.  
In the  latter, a sizeable initial axion velocity is the source of the axion relic abundance as opposed to the conventionally assumed initial misalignment angle. 
The early stage of oscillations in the $Z_\N$ axion framework naturally flows out into  kinetic misalignment.
 \begin{figure}[ht]
\centering
\includegraphics[width=0.5\textwidth]{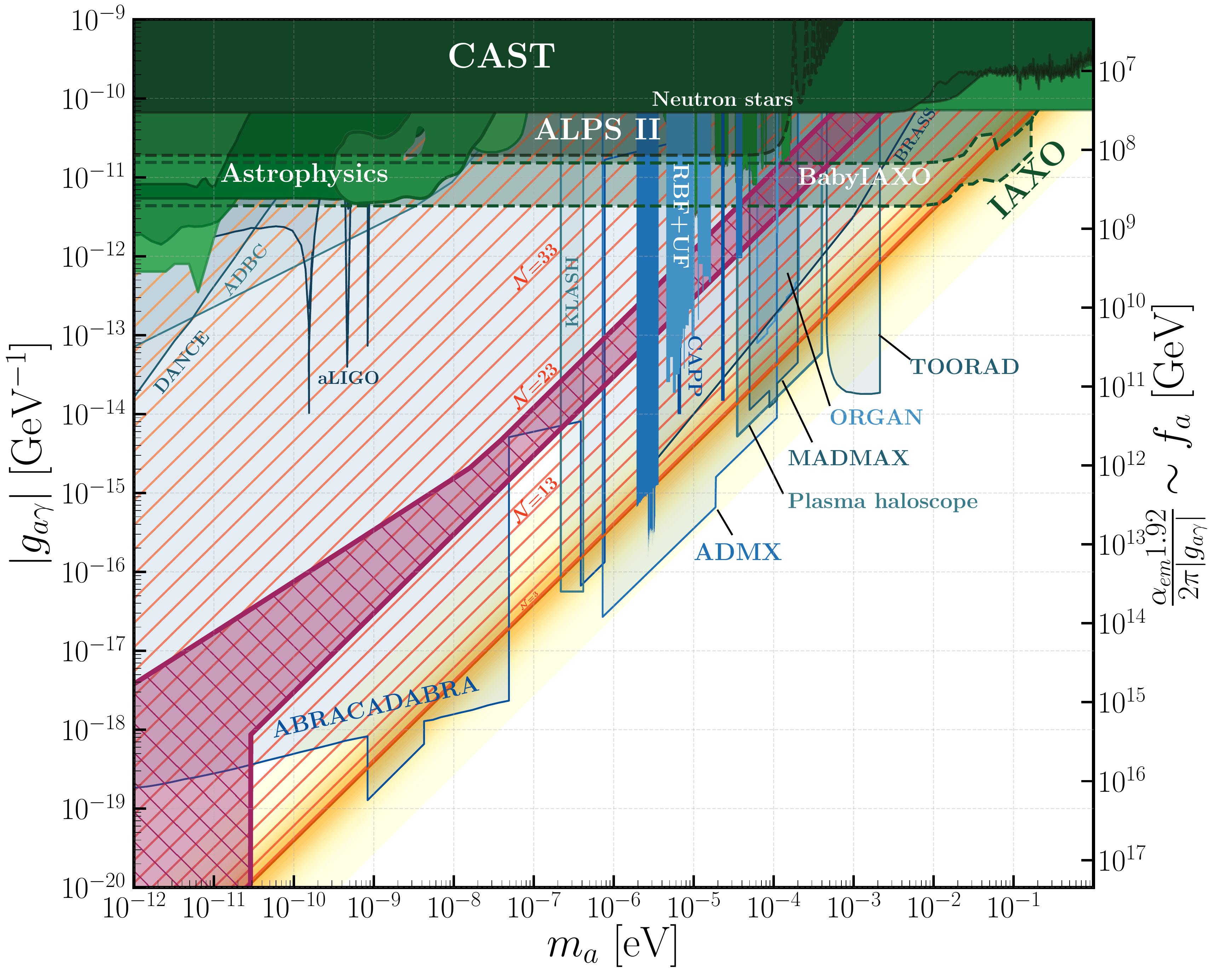} 
\includegraphics[width=0.45\textwidth]{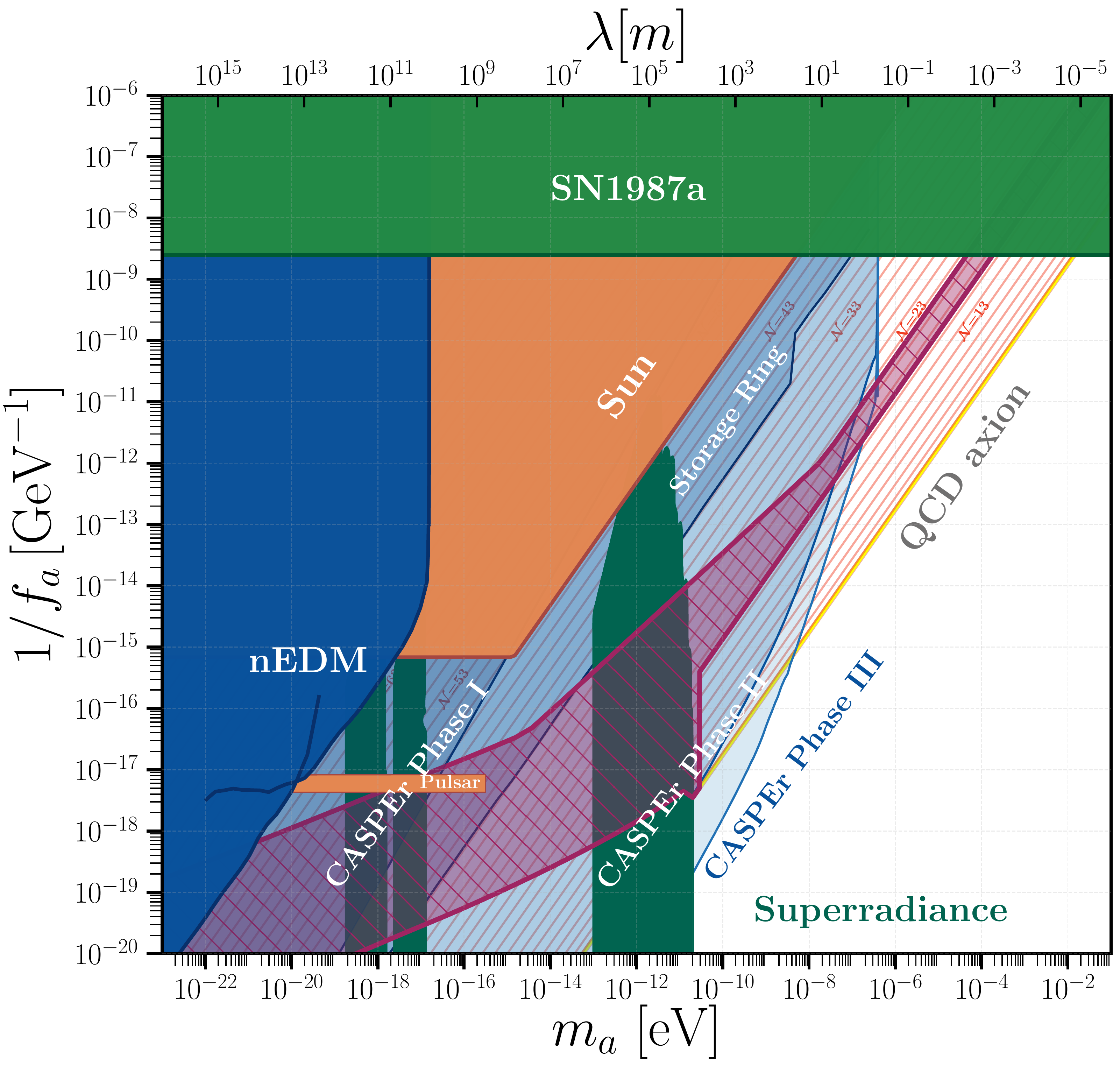}
\caption{  Limits on the axion-photon coupling (left) and on the inverse of the axion decay constant (right) as a function of the axion mass, assuming the axion accounts for the entire DM relic density. See Refs.~\cite{ciaran_o_hare_2020_3932430,DiLuzio:2021gos}.}
\label{fig:axionEDM}       
\end{figure}  
 \vspace*{-6pt}

The interplay of the different mechanisms together with the implications of the $Z_\N$ reduced-mass axion  
for axion DM searches is studied in detail in \cite{DiLuzio:2021gos}, including the experimental 
prospects to probe its coupling to photons, 
nucleons, electrons and the nEDM operator. As an example the bounds and projections of the $Z_\N$ axion assuming it accounts for all the DM abundance are depicted in Fig.~(\ref{fig:axionEDM}) for the coupling to photons and to gluons (i.e. $1/f_a$). 
The {\color{purpleDMbands}\bf purple} band in Fig.~(\ref{fig:axionEDM}) encompasses the region where the prediction of the $Z_\N$ axion relic density within the different regimes of the trapped misalignment can account for the entire DM abundance.
 As a wonderful byproduct of the lower-than-usual $f_a$ values 
 allowed in the $Z_\N$ 
 axion paradigm to solve the strong CP problem, 
  all axion-SM couplings are equally enhanced for a given $m_a$.  This increases the testability of the theory in current and future experiments, see Fig.~(\ref{fig:axionEDM}). It follows that the $Z_\N$ paradigm is --to our knowledge-- the only true axion theory  that could explain a positive signal in  CASPEr-Electric phase I and in a large region of the parameter space in phase II. 
 Moreover, the $Z_\N$ axion scenario includes --to our knowledge-- \emph{the first technically natural axion model of fuzzy DM that can also solve the strong CP problem}.
 \vspace*{-10pt}
{\section*{Acknowlegments}
\vspace*{-9pt}
  { \footnotesize P.Q. acknowledges support by the Deutsche Forschungsgemeinschaft (DFG, German Research Foundation) under Germany's Excellence Strategy -- EXC 2121 \textit{Quantum Universe} -- 390833306 and from the EU H2020 research and innovation programme under the Marie Sklodowska-Curie grant agreements 690575  (RISE InvisiblesPlus) and  674896 (ITN ELUSIVES) and 860881-HIDDeN, as well as from  the Spanish Research Agency through the grant IFT Centro de Excelencia Severo Ochoa SEV-2016-0597. 
 }
}

\bibliographystyle{utphys.bst}

\let\OLDthebibliography\thebibliography
\renewcommand\thebibliography[1]{
  \OLDthebibliography{#1}
  \setlength{\parskip}{1pt}
  \setlength{\itemsep}{3pt plus 0.3ex}
}

{\footnotesize  \setstretch{1} 
 

\providecommand{\href}[2]{#2}\begingroup\raggedright\endgroup
}

\end{document}